\documentstyle[aps,prl,epsf]{revtex}
\def\a{\alpha}
\def\b{\beta}

\def\d{\delta}

\begin{document}
\draft

\twocolumn[\hsize\textwidth\columnwidth\hsize\csname 
@twocolumnfalse\endcsname

\title{How fast can fluids squeeze through micro-pores?}

\author{Tom Chou}

\address{DAMTP, Cambridge University, Cambridge CB3 9EW, ENGLAND}

\maketitle

\begin{abstract}

We use a one dimensional symmetric exclusion model to study pressure
and osmosis driven flows through molecular-sized channels, such as
biological membrane channels and zeolite pores.  Analytic expressions
are found for the steady-state flow which show rich nonlinear behavior.
We find a maximum in the flow rates depending upon pore radius, pore
energetics, reservoir temperature, and driving force.  We also present
exact mean-field results of transport through pores with internal
defects. The interesting nonlinear dependences suggest numerous
diagnostic experiments in biological and zeolitic systems which may
reveal the features presented.  

\end{abstract}
\vspace{5mm}
]

The answer to the title question is tremendously important in analyzing
biological and industrial processes, and has received recent attention
with the finding that tracer particle motions in a micro-pore are
governed by subdiffusive dynamics \cite{KARGER}.  Biological examples
include membranes containing molecular-sized channels specific to water
transport which participate in cellular volume regulation controlled by
hydrostatic or osmotic pressure \cite{ALBERTS,FINK}. Classes of
biological ion channels have also been demonstrated to be pores of
molecular size \cite{ALBERTS,FINK}.  Man-made materials such as
zeolites and carbon nanotubes also contain many microscopic,
statistically nearly single-file channels which can selectively absorb
fluids.  This size specificity can be exploited in separation of a
mixture of linear and branched chain alkanes, where the zeolite acts as
like a sponge absorbing only the desired specie(s) \cite{BARRER}.
Confining particles in zeolite pores can also serve to catalyze
reactions: How fast can one get reagents into micro-pores and the
products out? Therefore, the design and manufacture of porous materials
is an economically motivated area of research.  Not surprisingly, vast
amounts of numerical simulation have been performed on a
variety of specific systems \cite{SHITMULATIONS}.

Confined particles are strongly interacting due to excluded
volume \cite{KARGER,CHOU95} as shown in recent NMR and
theoretical studies \cite{KARGER}.  
Anomalies in numerically computed
(MD) ``diffusion'' constants have been found
\cite{INDIAN}.  However, numerical simulations neither
access the long time scales required to study steady state
flow, nor offer a unifying physical picture of the
parameter regimes important for transport.  To obtain
reasonable flow rates using MD, artificial external forces
such as gravity are often imposed\cite{FLOWSHIT,CRACKNELL};
in 1D systems such external forces can yield qualitatively
different behavior (such as shock profiles) from osmosis
and pressure driven flow \cite{ASEM}, which occur in the
absence of such intrinsic forces.  

\begin{figure}[htb]
\begin{center}
\leavevmode
\epsfysize=2.6in
\epsfbox{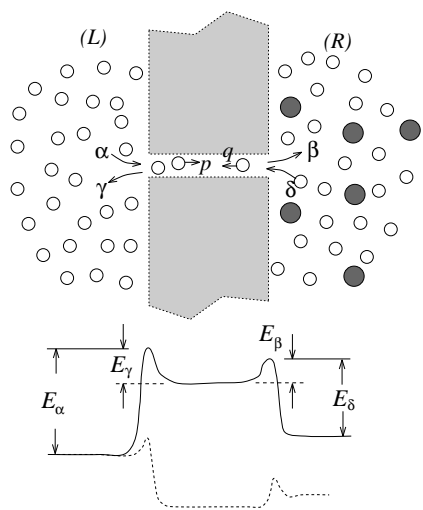}
\end{center}
\caption{Schematic of osmosis and pressure driven flow through membrane pores separating
infinite reservoirs $(L)$ and $(R)$.  The coefficients 
$\alpha,\beta,\gamma,\delta$ are conditional solvent entrance and exit 
probabilities at pore ends.}
\label{FIGURE1}
\end{figure}

A model of micro-flow that physically describes microscopic transport and
how it depends on macroscopic experimental parameters would serve as a
useful benchmark in more sophisticated models and complement more detailed
MD simulations.  Consider the molecular-sized pore shown in Fig. 
\ref{FIGURE1}, with particles driven from $(L)$ to $(R)$ either by osmotic
``pressure'' $\Delta\Pi$, or by hydrostatic pressure $\Delta P$.  
The pore is divided into sections $i$ of length $\ell$. 
Enthalpic energy differences between bulk and sectioned pore particles are
shown, as well as possible activation barriers.  Entrance(exit) rates at
the left and right boundary sites are denoted by $\alpha(\gamma)$ and
$\delta(\beta)$ respectively.  Here, $\alpha dt$ and $\delta dt$ are the
probabilities for pore entry in time $dt$ {\it only if} the occupations
$\tau_{i=1}=0$ and $\tau_{i=N}=0$ respectively. The probability per unit
time a randomly picked interior particle moves one section to the
right(left) is $p(q)$ {\it only if} the site to the right(left) is
unoccupied. 

\noindent The assumption of no pass pores is accurate even for pores with
diameters few times that of the particles: Overtaking requires a restricted
subset of geometries and will be statistically rare.  Even when overtaking
occurs, interchange between particles at sites $i$ and $i+1$ does not
contribute to net flux.  Although our treatment is rigorous for no pass
pores, it is accurate as long as the average number of particles in a
cylindrical slice of length $\ell$ is $\lesssim 1$.  Consider the
instantaneous number flux between sections $i$ and $i+1$, 

\begin{equation}
\begin{array}{l}
J(t) = p\tau_{i}(t)(1-\tau_{i+1}(t))-q\tau_{i+1}(t)(1-\tau_{i}(t)) \\[13pt]
\hspace{5mm} = p(\tau_{i}(t)-\tau_{i+1}(t))
+\epsilon\tau_{i+1}(t)-\epsilon\tau_{i}(t)\tau_{i+1}(t).
\label{JINSTANT}
\end{array}
\end{equation}

\noindent Although much attention has focussed on the asymmetric exclusion
model, particularly in the thermodynamic limit, where many exact results
are known \cite{ASEM}, the symmetric exclusion model is valid in the
absence of external electric or gravitational forces,  $\epsilon = p-q =
0$.  This reveals that hydraulic and osmotic pressure driven flows through
microscopic pores are simply a consequence of a density gradient which
induces particles to directionally or cooperatively diffuse; however,
locally, a particle is as likely to move to the left or right {\it if} both
left and right adjacent sites are unoccupied. The random updating in
particle motions implicit in the model dynamics is clearly valid in systems
where long wavelength collective modes are irrelevant.  


Linearity of $J(t)$ when $\epsilon = 0$ renders the
mean-field steady state current $J$ found from time averaging
Eq. (\ref{JINSTANT}) exact.  Steady state particle conservation along the
pore length results in a linear density profile  $J =
p(\tau_{1}-\tau_{N})/(N-1)$; this, along with the steady state boundary
conditions $J_{1}=J=\alpha(1-\tau_{1})-\gamma\tau_{1}$ and
$J_{N}=J=\beta\tau_{N}-\delta(1-\tau_{N})$, determine the steady state
particle number flux

\begin{equation}
J(N) = {p(\alpha\beta - \gamma\delta) \over
(N-1)(\alpha+\gamma)(\beta+\delta) + p(\alpha+\beta+\gamma+\delta)}.
\end{equation}

The kinetic parameters $\{\mu\} \equiv
(\alpha,\beta,\gamma,\delta)$ are related to the
{\it relative} enthalpies of activation $E_{\mu}$
between pore and bath particles. We assume local
thermodynamic equilibrium (LTE), particularly
valid in liquid phase osmosis across single
biological pores where $J \leq 10^{9}/$s, typical
pore diameters and interparticle spacings
$\lambda \sim 5\mbox{\AA}$, and ambient thermal
velocities $v_{T} \simeq 4\times 10^{4}$cm/s,
yield collision times $t_{coll}\simeq
\lambda/v_{T} \simeq 1 \mbox{ps} \ll J^{-1}$. 
Therefore, particles suffer $O(10^{3})$
collisions before they are osmotically
transported, sufficient for (LTE).  As an
illustrative example, we consider an axisymmetric
right cylindrical pore.  The kinetic parameters
under LTE will be defined by simple transport
theory, 

\begin{equation}
\begin{array}{l}
p \approx (v_{T}/\ell)\exp(-E_{p}/k_{B}T);\,\,
\beta \approx \left( v_{T}/\ell \right)
\exp(-E_{\beta}/k_{B}T)\\[13pt]
\alpha \equiv \alpha_{0}\exp(-E_{\alpha}/k_{B}T)  \approx {1\over 4}n_{L}v_{T} (\pi
r_{p}^{2}) \exp(-E_{\alpha}/k_{B}T)
\label{PARAMETERS}
\end{array}
\end{equation}

\noindent where $v_{T}\sim \sqrt{k_{B}T/m}$ is the thermal velocity, $\ell$
is chosen to be the minimum statistical spacing between pore particles, and
$n_{L}$ is the number density in the left reservoir. The minimum particle
spacing $\ell$ can be estimated for each set of $\{\mu\}$ from a 1D Tonk's
gas \cite{TONKS}.  Since we explore considerable variations in $\{\mu\}$,
we choose $\ell \sim a$, approximately a hard core repulsive diameter;
thus, $p$ given in (\ref{PARAMETERS}) represents a ballistic travel time
over the distance $\ell \simeq a$, weighted by an energetic binding
$E_{p}$. A larger choice for $\ell$ can be made if $(\ell/a)\tau_{i} < 1$
and with $p$ appropriately rescaled and entropic factors included for
$\{\mu\}$ (the $E_{\mu}$ are then effective free energies); this is useful
in multiple species models where steady state flows for long pores cannot
be obtained analytically \cite{PRE}.  Note that for $\ell \gg a$, {\it
local} free diffusive transport described by $p\simeq \ell^{-2}$ may
obtain. But when $\ell \simeq a$, $E_{\mu}$ represent enthalpies determined
entirely by molecular potentials.  Thus, $E_{\alpha}(r\leq
r_{p})-E_{\alpha}(0) \lesssim k_{B}T$ (where $E_{\alpha}(r=0) \equiv
E_{\alpha}$) defines an effective pore radius $r_{p}$. For pores that repel
particles (top curve in Fig. \ref{FIGURE1}) and have negligible activation
energies $E_{\beta}$ and $E_{p}$, $p/\b \sim O(1)$.  

Further simplification is gained by considering microscopically symmetric
pores where $\beta=\gamma$, and normalizing (denoted by an overbar) all
quantities by $\beta$ (such that $\bar{J} = 1$ would be the maximum flow
rate possible, when $\tau_{N}=1$) so that

\begin{equation}
\bar{J}(N)={\bar{\alpha}\bar{p}\Delta \over (N-1)(\bar{\alpha}+1)(\bar{\alpha}+1-
\bar{\alpha}\Delta) + \bar{p}(2\bar{\alpha}+2-\bar{\alpha}\Delta)},
\label{JN}
\end{equation}

\noindent  where

\begin{equation}
\Delta = 1-e^{\Delta E/k_{B}T}+ {\Delta n_{solute}
\over n_{L}}e^{\Delta E/k_{B}T}
\label{DELTA}
\end{equation}

\noindent and $\Delta E \equiv E_{\a}-E_{\d}(P_{R}-P_{L})$. 
Eqn.  (\ref{DELTA}) represent differences in number density
and/or enthalpies between $(R)$ and $(L)$ and along with
(\ref{JN}) determine the flow through symmetric pores.  In
osmotic flow under isobaric conditions, $\Delta E = 0$, in
pressure driven flow of nearly ideal gases, $(\partial \Delta
E /\partial P_{R})_{T} \simeq 0$, while pressure driven flows
of liquids is described by $(\partial n_{R}/\partial P)_{T}
\simeq 0$.  In the first two cases, the flux is predominately
the consequence of increased permeable particle density in one
of the reservoirs over the other, while pressure driven flow
of liquids results mainly from the relative reduction of pore
entrance activation energies  brought about by hydrostatic
compression. We first consider fixed $\Delta = 0.02$ which
corresponds to an osmotic pressure in aqueous solution of
$\Delta\Pi \simeq 25$ atm or a hydrostatic pressure difference
of $\Delta P \simeq 0.025$ atm of gas at STP. Using the
Maxwell relationship for molar volume, $-(\partial \Delta
E/\partial P_{R})_{T} \simeq \tilde{v}$, we find $\Delta =
0.02$ also corresponds to $P_{R}-P_{L} \simeq 25$atm in
pressure driven flow of water at 300K.

When $\bar{\alpha}\Delta/(\bar{\alpha}+1)$ is negligible, flows are
essentially linear in $\Delta$ and defined by hydraulic or osmotic
permeabilities, $J = L_{p}\Delta P$ or $J=P_{os}\Delta\Pi$. 
Although $L_{p}$ and
$P_{os}$ have often been, and continue to be interpreted using macroscopic
fluid mechanics \cite{FINK}, 
a microscopic description arises here. In the limit $\bar{p} \gg
(\bar{\alpha}+1)N$, rate limiting steps involve pore entrance or exit. 
Linearizing (\ref{DELTA}) and (\ref{JN}), we find 

\begin{equation}
L_{p} = {\a \over 2(\bar{\a}+1) k_{B}T}\left({\partial \Delta E \over
\partial P_{R}}\right)_{T}
\label{LP1}
\end{equation}

\noindent for pressure driven flow of dense liquids.  The temperature
dependence of $L_{p}$ will be determined by
$-E_{\beta}/k_{B}T\,\left[-E_{\alpha}/k_{B}T\right]$ for $\bar{\alpha} \gg
1\,\left[\bar{\alpha} \ll 1 \right]$ if the thermal coefficient of
expansion $\approx 0$.  When $\bar{p} \ll (\bar{\alpha}+1)N$,


\begin{equation}
L_{p} = {\bar{\a} \bar{q} \over (N-1)(\bar{\a}+1)^{2}k_{B}T}\left(
{\partial \Delta E \over
\partial P_{R}}\right)_{T}
\label{LP2}
\end{equation}

\noindent which has a
$(E_{\alpha}-E_{\beta}-E_{p})/k_{B}T\,
\left[(E_{\beta}-E_{\alpha}-E_{p})/k_{B}T\right]$
Arrhenius temperature dependence for
$\bar{\alpha} \gg 1\,\left[\bar{\alpha}\ll
1\right]$.  Note that the $\bar{\a}\gg 1$ regime
of (\ref{LP2}) yields a curious $L_{p} \propto
r_{p}^{-2}$ dependence. In the limit where
(\ref{LP2}) holds, the bottlenecks occur in the
particle motions within the pore interior. 
Expressions for $L_{p}$ of ideal gases and
$P_{os}(P_{L}=P_{R})$ are found from (\ref{LP1})
and (\ref{LP2}) by replacing $(\partial \Delta E
/\partial P_{R})_{T} \rightarrow \mp n_{L}^{-1}$
respectively in the corresponding limits; the
temperature dependences remain unchanged. Values
of all energetic barriers mentioned above are
qualitatively consistent with osmotic flow
through biological pores where hydrogen bonds
($E_{\alpha}\sim 10k_{B}T$ in the
$\bar{\alpha}\ll 1$ limit of (\ref{LP2})) must be
broken before water can enter.


\begin{figure}[htb]
\begin{center}
\leavevmode
\epsfysize=2.7in
\epsfbox{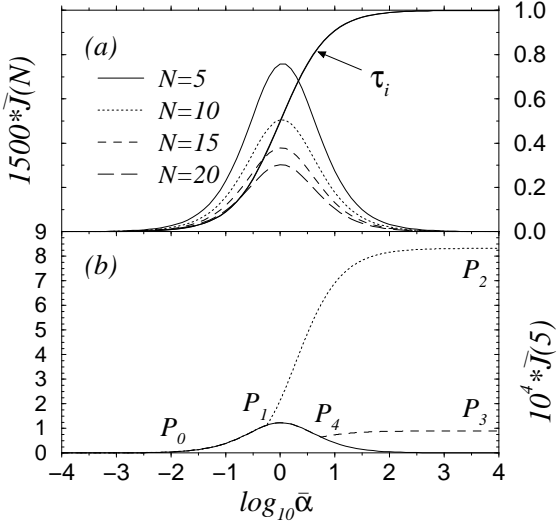}
\end{center}
\caption{(a) $1.5\times 10^{3}\bar{J}(N,\bar{\alpha})$ for $\Delta = 0.02$ and 
fixed $\bar{p}_{0} = 1.0$. Various lengths are indicated. The 
solid curve between 0 and 1 is the average occupation number of 
any site. On this scale, the difference $\tau_{1}-\tau_{N}$ 
is not apparent but varies qualitatively as $\bar{J}$. (b) 
The solid curve is $10^{4}\times \bar{J}(5)$ for fixed $\bar{p}_{0}=0.1$
See text for explanation of 
curves $P_{0}P_{1}P_{2}$ and $P_{0}P_{1}P_{4}P_{3}$.}
\label{FIGURE2}
\end{figure}

The solid curve in Figure \ref{FIGURE2}(a) shows $(1.5\times
10^{3})\times\bar{J}(N)$ (from Eq. (\ref{JN})) for $\bar{p}=1,\, \Delta =
0.02$ for various pore lengths $N\simeq L/a$. We find a maximum flux
$\bar{J}^{*}$ at 

\begin{equation}
\bar{\alpha}^{*} = \left[{2\bar{p}_{0}+(N-1) \over (N-1)(1-\Delta)}\right]^{1/2}
\label{ALPHAMAX}
\end{equation}

\noindent for fixed $\bar{p}=\bar{p}_{0}$. The maximum occurs because at
intermediate occupancies, the pore is conducting a significant number of
particles, but is not yet choked off. Large values of $\bar{\alpha}$
represent pores which are attractive to the solvent, depicted by the dashed
energy landscape in Fig.  \ref{FIGURE1}.  As the pore is made increasingly
attractive $E_{\beta} > 0$, and  $\beta$ must eventually diminish so that
$\bar{p}=p/\b\propto \exp\left[(E_{\beta}-E_{p})/k_{B}T\right]$.  However,
we will explicitly show that the maximum at $\bar{J}^{*}$ can persist. 
Assume no activation barriers at the pore mouths, {\it ie.,}
$E_{\beta}(E_{\alpha}) = 0$ for $E_{\alpha}(E_{\beta}) > 0$ and consider
molecularly repelling pores where $0< \bar{\alpha}<\bar{\alpha}_{0}$
($E_{\alpha}>0,\, E_{\b}=0$); here, $\bar{p} \approx \exp(-E_{p}/k_{B}T)$,
independent of the pore energy level.  As the pore is made attracting,
$\bar{p}$ will acquire $\exp(E_{\b}/k_{B}T)$ behavior and can be defined as

\begin{equation}
\bar{p}=\bar{p}_{0}\left[{\bar{\alpha}\over
\bar{\alpha}_{0}}\theta(\bar{\alpha}-\bar{\alpha}_{0})+
\theta(\bar{\alpha}_{0}-\bar{\alpha})\right].
\label{PTHETA}
\end{equation}

\noindent where $\theta(x>1) = 1$ is the Heaviside function indicating when
the pore first becomes molecularly attracting.  Using (\ref{PTHETA}),
$\bar{J}(N,\bar{\a}\rightarrow \infty)$ (the current in the limit of
infinitely attracting pores)  becomes

\begin{equation} 
\bar{J}^{\infty}(N) = {\bar{p}_{0} \Delta \over
\bar{\alpha}_{0}(N-1)(1-\Delta)+\bar{p}_{0}(2-\Delta)},
\label{JINFTY} 
\end{equation}

\noindent which can approach $\Delta/(2-\Delta) > \bar{J}^{*}$.  For
$\bar{\alpha}_{0} \gg \bar{\alpha}^{*}$, the maximum remains at
$\bar{\alpha}^{*}$.  However, when $\bar{\alpha}_{0} \lesssim
\bar{\alpha}^{*},\, \bar{\alpha}$, $\bar{p} \approx \exp(E_{\beta}/k_{B}T)$
and the maximum in flux as a function of $\bar{\a}$ is preempted by a
current which approaches (\ref{JINFTY}). The condition for
$\bar{J}^{\infty}(N) < \bar{J}^{*}(N)$ (a remaining maximum in $\bar{J}$ as
pore well depth is increased) is determined by

\begin{equation}
\bar{\alpha}_{0} > \bar{\alpha}^{*}+{\bar{\alpha}^{*2}\over \bar{\alpha}^{*}+1}
\label{CONDITION}
\end{equation}
 
\noindent Fig. \ref{FIGURE2}(b) compares the behavior of $10^{4}\times J(5)$
using $\bar{p}=\bar{p}_{0}=0.1$ (solid curve) with that of $10^{4}\times
J(5)$ using Eq. (\ref{PTHETA}) (broken lines).  For $\Delta=0.02$, the
maximum at $\bar{\alpha}^{*}=\sqrt{15/14}$ is destroyed when
(\ref{CONDITION}) is satisfied, $\bar{\alpha}_{0} \lesssim 1.562$. 
Estimating $\bar{\alpha}_{0}$ from (\ref{PARAMETERS}) and $\Delta = 0.02$,
$\bar{\alpha}_{0} \ll \bar{\alpha}^{*}$ for gases, but can be $O(1)$ for
liquids. Curve $P_{0}P_{1}P_{4}P_{3}$ retains the maximum $\bar{J}^{*}$
since $\bar{\alpha}_{0} = 10^{0.75} > 1.562$, while $P_{0}P_{1}P_{2}$
corresponds to $\bar{\alpha}_{0} = 10^{-0.25} < 1.562$ which gives a
monotonic $J$ as $\bar{\alpha}\rightarrow \infty$.  A high current may
occur at $\bar{\alpha}\rightarrow \infty$ despite the high pore occupancy
due to an accompanying exponential increase in $\bar{p}$.  As expected,
these high values are more difficult to achieve as $\bar{\alpha}_{0}$ (as
well as $N$) {\it increases}, because the onset of exponentially
increasing $\bar{p}$ is delayed.  

\begin{figure}[htb] 
\begin{center} 
\leavevmode 
\epsfysize=2.2in
\epsfbox{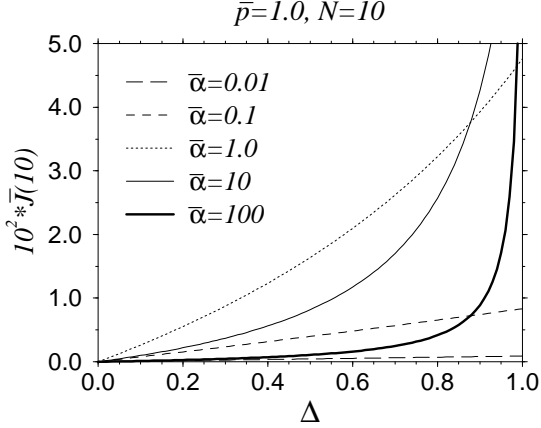} 
\end{center} 
\caption{Nonlinear behavior of $10^{2}\times
\bar{J}(10)$ for $\bar{p}=1.0$ as a function of $\bar{\alpha}$. Note the
competition between nonlinearities in $\bar{\a}$ and $\Delta$, particularly at
large $\Delta$. } 
\label{FIGURE3} 
\end{figure}

Now consider the nonlinear effects of large $\Delta$.  According to
(\ref{ALPHAMAX}), values of $\bar{\alpha}$ that give a maximal
$\bar{J}^{*}$ depend strongly on $\Delta$; thus, $\Delta \rightarrow 1$ can
yield large $\bar{\alpha}^{*}\gg \bar{\alpha}_{0}$ where the maximum in
flux is destroyed. At the maximum value $\Delta=1$ (corresponding to
approximately to pure solute or vacuum in $(R)$), the maximum flux occurs
when $\bar{p}\gg N$ and approaches $\bar{J}(\Delta=1)\simeq
\bar{\alpha}/(\bar{\alpha}+2)$.  The smallest flux occurs when $\bar{p}\ll
N$ and $\bar{\alpha} \ll 1$ as expected.  The nonlinearity of
$\bar{J}(\Delta)$ is important only when $\bar{\alpha}\gg 1$, as shown in 
Figure \ref{FIGURE3}, corresponding to a pore interior with high particle
occupation, when particle exclusion nonidealities are most pronounced.

\noindent To experimentally test possible nonlinearities
densities, pressures, and temperatures
may need to be substantially varied, and is readily performed 
only in 
zeolite systems. Adiabatic ultrasonic
driving with frequency $\omega \ll J^{-1}$ of the fluid in
$(R)$ (not affecting $\bar{\alpha}$ which depends only on
$n_{L}$) will enhance transport: Upon setting $\Delta(t) =
\Delta_{0}+\Delta_{1}\cos\omega t$ and averaging over a
period, 


\begin{equation}
\begin{array}{l}
\displaystyle \bar{J}(N) = \bar{J}(N,\Delta_{1}=0) + 
{\bar{J}(N,\Delta_{1}=0)\over 2}\times \\[13pt]
\hspace{9mm}\left\{
\begin{array}{ll}
{\bar{\alpha}^{2}\Delta_{1}^{2}\over \bar{\alpha}(2-\Delta_{0})+1} &
\bar{p}\gg(\bar{\alpha}+1)N \\[13pt]
{\Delta_{1}^{2}\over (1-\Delta_{0})^{2}} & \bar{p}\ll (\bar{\alpha}+1)N
\end{array} \right\}+O(\Delta_{1})^{3}
\end{array}
\end{equation}

\noindent for small $\Delta_{1}/\delta_{0}$. In the $\bar{p}\gg (\bar{\a}+1)N$ 
case, the temperature dependence of the flux enhancement is
$-2E_{\alpha}/k_{B}T$ or $-E_{\alpha}/k_{B}T$, while the second case is
temperature independent.


Effects of microscopic internal pore structure/disorder on microflow can
also be modelled.  Consider $N^{*}$ defects in hopping rates $p^{*}_{k}(t)
= q^{*}_{k}(t)$, between sections $i(k)$ and $i(k)+1$. When $p^{*}_{k}(t)$
is statistically independent of $\tau_{i}(t)$, the mean-field result 

\begin{equation}
J = {p(\alpha\beta-\gamma\delta) \over (\alpha+\gamma)(\beta+\delta)
(N-1+\Omega)+p(\alpha+\beta+\gamma+\delta)},
\label{JDEFECT}
\end{equation}

\noindent where $\Omega\equiv \sum_{k=1}^{N^{*}}(p/p^{*}_{k}-1)$, is exact. 
Biological and zeolitic channels contain internal pore structure exposing random
binding sites $E_{p}(k)$. An example of a one-defect pore is gramicidin A
channels across a biomembrane.  These are composed of two barrel structures, one
in each lipid layer, joined by attractive molecular interactions near the bilayer
midplane forming a defect bisecting the entire channel. Osmosis experiments on
Gramicidin A/bilayer liposomes reveal rich temperature dependences which are
interpreted as lipid phase transitions inducing changes in how the Gramicidin A
barrels are joined \cite{BOEHLER}.  Molecular permeation through pure lipid
bilayers is also modelled by (\ref{JDEFECT}).  Biolipid headgroups offer
activation barriers such as those depicted in Fig \ref{FIGURE1}. 
Within the hydrophobic lipid tails, particle motions are predominantly
perpendicular to the membrane along the aliphatic chains. Thus, bilayer membranes
physically resemble many close-packed channels with defects at the bilayer
midplane, where the ends of the hydrocarbon chains approach each other, and at
stiff unsaturated bonds along the fatty acid chains. Various zeolites are
comprised of interconnected cages and joints which are also modelled using
(\ref{JDEFECT}), although  actual flow measurements are ensemble averages over
macroscopic membrane regions containing many pores: $\langle \bar{J}\rangle =
\sum_{N,N^{*},p^{*}}^{\infty} f(N,N^{*},p_{k}^{*}) \bar{J}(N,N^{*},p_{k}^{*})$,
where $d$ is the membrane thickness, $N_{\em min} \simeq d/\ell$, and
$f(N,N^{*},p_{k}^{*})$ is the distribution of channels with arc-length
$L=N/\ell$, number of defects $N^{*}$, and defect strengths $p_{k}^{*}$.  We
mention that the steady state result (\ref{JDEFECT}) is also an exact solution to
the Heisenberg spin chain hamiltonian with boundary conditions determined by
$\{\mu\}$ and quenched random energies \cite{SANDOW}.


Since the particle density at section $i$ obeys $\dot{\tau}_{i} =
q(\tau_{i+1}-2\tau_{i}+\tau_{i-1})$ which on scales $\gg \ell$ is a
diffusion equation with $q\ell ^{2}$ defining a cooperative diffusion
coefficient, osmosis occurs strictly by mass diffusion, different from
tracer diffusion.  Conditions required for a maximum in $J(\bar{\a})$
depend on the pre-exponential factor $\alpha_{0}$ for realistically defined
$\{\mu\}$. In LTE, relationships among the various kinetic parameters and
various fluids can be determined by equilibrium measurements such as
solvent-solute volumes and heats of mixing.  The different Arrhenius
temperature regimes delineated may also be simple way of systematically
probing solvent-pore interactions.  Experimentally the nonlinearities in
$\Delta$ may be probed by low frequency acoustic driving.  Electrostriction
and mechanical deformation of the pores may also control $\bar{\a}$ and $p$
via $r_{p}$ \cite{INDIAN}. Effects of unstirred layers near the pore mouths
can be easily treated with macroscopic convection-diffusion equations, but
are experimentally \cite{FINK} and theoretically \cite{PRE} found to be
unimportant in aqueous osmosis.  When important, unstirred layers yield an
implicit equation $J\propto \Delta(J)$ \cite{PRE}. The simple model
presented along with the numerous applicable physical systems and proposed
experimental tests complements MD simulations and will subsequently lead to
a better understanding of more complex systems, including multispecies
transport and chemical reactions in pores.

\end{document}